

\input amssym.tex 

\def\unredoffs{}
\tolerance=1000\hfuzz=2pt
\catcode`\@=11 
\ifx\hyperdef\UNd@FiNeD\def\hyperdef#1#2#3#4{#4}\def\hyperref#1#2#3#4{#4}\def\href#1#2{#2}\fi
\magnification=1200\unredoffs\baselineskip=16pt plus 2pt minus 1pt
\def\Date#1{\vfill\leftline{#1}\tenpoint\supereject%
\footline={\hss\tenrm\hyperdef\hypernoname{page}\folio\folio\hss}}%

{\count255=\time\divide\count255 by 60 \xdef\hourmin{\number\count255}
 \multiply\count255 by-60\advance\count255 by\time
 \xdef\hourmin{\hourmin:\ifnum\count255<10 0\fi\the\count255}
}
\def\date{\number\day.\number\month.\number\year\ at \hourmin}


\def\nolabels{\def\wrlabeL##1{}\def\eqlabeL##1{}\def\reflabeL##1{}}
\def\writelabels{\def\wrlabeL##1{\leavevmode\vadjust{\rlap{\smash%
{\line{{\escapechar=` \hfill\rlap{\sevenrm\hskip.03in\string##1}}}}}}}%
\def\eqlabeL##1{{\escapechar-1\rlap{\sevenrm\hskip.05in\string##1}}}%
\def\reflabeL##1{\noexpand\llap{\noexpand\sevenrm\string\string\string##1}}}
\nolabels

\global\newcount\secno \global\secno=0
\global\newcount\meqno \global\meqno=1
\def\s@csym{}

\def\newsec#1\par{\global\advance\secno by1%
{\toks0{#1}\message{(\the\secno. \the\toks0)}}%
\global\subsecno=0\eqnres@t\let\s@csym\secsym\xdef\secn@m{\the\secno}\noindent
{\bf\hyperdef\hypernoname{section}{\the\secno}{\the\secno.} #1}%
\writetoca{{\string\hyperref{}{section}{\the\secno}{\bf \the\secno\quad}} {\bf #1}}\par%
\nobreak\medskip\nobreak\noindent\ignorespaces}
\def\eqnres@t{\xdef\secsym{\the\secno.}\global\meqno=1\bigbreak\bigskip}
\def\sequentialequations{\def\eqnres@t{\bigbreak}}\xdef\secsym{}

\global\newcount\subsecno \global\subsecno=0
\def\subsec#1\par{\global\advance\subsecno by1%
{\toks0{#1}\message{(\s@csym\the\subsecno. \the\toks0)}}%
\global\subsubsecno=0%
\ifnum\lastpenalty>9000\else\bigbreak\fi
\noindent{\it\hyperdef\hypernoname{subsection}{\secn@m.\the\subsecno}%
{\secn@m.\the\subsecno.} #1}\writetoca{\string\hskip1.45cm
{\string\hyperref{}{subsection}{\secn@m.\the\subsecno}{\secn@m.\the\subsecno.}}
{#1}}\par\nobreak\medskip\nobreak\noindent\ignorespaces}

\def\appendix#1#2{\global\meqno=1\global\subsecno=0\xdef\secsym{\hbox{#1.}}%
\bigbreak\bigskip\noindent{\bf Appendix \hyperdef\hypernoname{appendix}{#1}%
{#1.} #2}{\toks0{(#1. #2)}\message{\the\toks0}}%
\xdef\s@csym{#1.}\xdef\secn@m{#1}%
\writetoca{{\string\hyperref{}{appendix}{#1}{\bf {#1}\quad}} {\bf #2}}%
\par\nobreak\medskip\nobreak}

%
\def\checkm@de#1#2{\ifmmode{\def\f@rst##1{##1}\hyperdef\hypernoname{equation}%
{#1}{#2}}\else\hyperref{}{equation}{#1}{#2}\fi}
\def\eqnn#1{\DefWarn#1\xdef #1{(\noexpand\relax\noexpand\checkm@de%
{\s@csym\the\meqno}{\secsym\the\meqno})}%
\wrlabeL#1\writedef{#1\leftbracket#1}\global\advance\meqno by1}
\def\f@rst#1{\c@t#1a\em@ark}\def\c@t#1#2\em@ark{#1}
\def\eqna#1{\DefWarn#1\wrlabeL{#1$\{\}$}%
\xdef #1##1{(\noexpand\relax\noexpand\checkm@de%
{\s@csym\the\meqno\noexpand\f@rst{##1}1}{\hbox{$\secsym\the\meqno##1$}})}
\writedef{#1\numbersign1\leftbracket#1{\numbersign1}}\global\advance\meqno by1}
\def\eqn#1#2{\DefWarn#1%
\xdef #1{(\noexpand\hyperref{}{equation}{\s@csym\the\meqno}%
{\secsym\the\meqno})}$$#2\eqno(\hyperdef\hypernoname{equation}%
{\s@csym\the\meqno}{\secsym\the\meqno})\eqlabeL#1$$%
\writedef{#1\leftbracket#1}\global\advance\meqno by1}
\def\xeqn{\expandafter\xe@n}\def\xe@n(#1){#1}
\def\xeqna#1{\expandafter\xe@n#1}
\def\eqns#1{(\e@ns #1{\hbox{}})}
\def\e@ns#1{\ifx\UNd@FiNeD#1\message{eqnlabel \string#1 is undefined.}%
\xdef#1{(?.?)}\fi{\let\hyperref=\relax\xdef\next{#1}}%
\ifx\next\em@rk\def\next{}\else%
\ifx\next#1\xeqn#1\else\def\n@xt{#1}\ifx\n@xt\next#1\else\xeqna#1\fi
\fi\let\next=\e@ns\fi\next}

\def\DefWarn#1{\ifx\UNd@FiNeD#1\else
\immediate\write16{*** WARNING: the label \string#1 is already defined ***}\fi}
%
\newskip\footskip\footskip14pt plus 1pt minus 1pt 
\def\footnotefont{\ninepoint}\def\f@t#1{\footnotefont #1\@foot}
\def\f@@t{\baselineskip\footskip\bgroup\footnotefont\aftergroup\@foot\let\next}
\setbox\strutbox=\hbox{\vrule height9.5pt depth4.5pt width0pt}
\global\newcount\ftno \global\ftno=0
\def\foot{\global\advance\ftno by1\def\foot@rg{\hyperref{}{footnote}%
{\the\ftno}{\the\ftno}\xdef\foot@rg{\noexpand\hyperdef\noexpand\hypernoname%
{footnote}{\the\ftno}{\the\ftno}}}\footnote{$^{\foot@rg}$}}
%
%
%
\global\newcount\refno \global\refno=1
\newwrite\rfile
\def\ref{[\hyperref{}{reference}{\the\refno}{\the\refno}]\nref}
\def\nref#1{\DefWarn#1%
\xdef#1{[\noexpand\hyperref{}{reference}{\the\refno}{\the\refno}]}%
\writedef{#1\leftbracket#1}%
\ifnum\refno=1\immediate\openout\rfile=\jobname.refs\fi
\chardef\wfile=\rfile\immediate\write\rfile{\noexpand\item{[\noexpand\hyperdef%
\noexpand\hypernoname{reference}{\the\refno}{\the\refno}]\ }%
\reflabeL{#1\hskip.31in}\pctsign}\global\advance\refno by1\findarg}
\def\findarg#1#{\begingroup\obeylines\newlinechar=`\^^M\pass@rg}
{\obeylines\gdef\pass@rg#1{\writ@line\relax #1^^M\hbox{}^^M}%
\gdef\writ@line#1^^M{\expandafter\toks0\expandafter{\striprel@x #1}%
\edef\next{\the\toks0}\ifx\next\em@rk\let\next=\endgroup\else\ifx\next\empty%
\else\immediate\write\wfile{\the\toks0}\fi\let\next=\writ@line\fi\next\relax}}
\def\striprel@x#1{} \def\em@rk{\hbox{}}
\def\lref{\begingroup\obeylines\lr@f}
\def\lr@f#1#2{\DefWarn#1\gdef#1{\let#1=\UNd@FiNeD\ref#1{#2}}\endgroup\unskip}
\def\semi{;\hfil\break}
\def\addref#1{\immediate\write\rfile{\noexpand\item{}#1}} 
\def\listrefs{\vfill\supereject\immediate\closeout\rfile\writestoppt
\baselineskip=\footskip\centerline{{\bf References}}\bigskip{\parindent=20pt%
\frenchspacing\escapechar=` \input \jobname.refs\vfill\eject}\nonfrenchspacing}
\def\startrefs#1{\immediate\openout\rfile=\jobname.refs\refno=#1}
\def\xref{\expandafter\xr@f}\def\xr@f[#1]{#1}
\def\refs#1{\count255=1[\r@fs #1{\hbox{}}]}
\def\r@fs#1{\ifx\UNd@FiNeD#1\message{reflabel \string#1 is undefined.}%
\nref#1{need to supply reference \string#1.}\fi%
\vphantom{\hphantom{#1}}{\let\hyperref=\relax\xdef\next{#1}}%
\ifx\next\em@rk\def\next{}%
\else\ifx\next#1\ifodd\count255\relax\xref#1\count255=0\fi%
\else#1\count255=1\fi\let\next=\r@fs\fi\next}
%

%
\newwrite\ffile\global\newcount\figno \global\figno=1
\def\fig{fig.~\hyperref{}{figure}{\the\figno}{\the\figno}\nfig}
\def\nfig#1{\DefWarn#1%
\xdef#1{fig.~\noexpand\hyperref{}{figure}{\the\figno}{\the\figno}}%
\writedef{#1\leftbracket fig.\noexpand~\xfig#1}%
\ifnum\figno=1\immediate\openout\ffile=\jobname.figs\fi\chardef\wfile=\ffile%
{\let\hyperref=\relax
\immediate\write\ffile{\noexpand\medskip\noexpand\item{Fig.\ %
\noexpand\hyperdef\noexpand\hypernoname{figure}{\the\figno}{\the\figno}. }
\reflabeL{#1\hskip.55in}\pctsign}}\global\advance\figno by1\findarg}
\def\xfig{\expandafter\xf@g}\def\xf@g fig.\penalty\@M\ {}
\def\figs#1{figs.~\f@gs #1{\hbox{}}}
\def\f@gs#1{{\let\hyperref=\relax\xdef\next{#1}}\ifx\next\em@rk\def\next{}\else
\ifx\next#1\xfig #1\else#1\fi\let\next=\f@gs\fi\next}
%
\def\figin{\epsfcheck\figin}\def\figins{\epsfcheck\figins}
\def\epsfcheck{\ifx\epsfbox\UnDeFiNeD
\message{(NO epsf.tex, FIGURES WILL BE IGNORED)}
\gdef\figin##1{\vskip2in}\gdef\figins##1{\hskip.5in}
\else\message{(FIGURES WILL BE INCLUDED)}%
\gdef\figin##1{##1}\gdef\figins##1{##1}\fi}
\def\DefWarn#1{}
\def\figinsert{\goodbreak\topinsert}
\def\ifig#1#2#3{\DefWarn#1\xdef#1{fig.~\the\figno}
\writedef{#1\leftbracket fig.\noexpand~\the\figno}%
\figinsert\figin{\centerline{#3}}
\smallskip
\leftskip=0pt \rightskip=0pt
\baselineskip12pt\noindent
{{\bf Fig.~\the\figno}\ \ninepoint #2}
\medskip
\global\advance\figno by1\par\endinsert}
\newwrite\lfile
{\escapechar-1\xdef\pctsign{\string\%}\xdef\leftbracket{\string\{}
\xdef\rightbracket{\string\}}\xdef\numbersign{\string\#}}
\def\writedefs{\immediate\openout\lfile=label.defs \def\writedef##1{%
{\let\hyperref=\relax\let\hyperdef=\relax\let\hypernoname=\relax
 \immediate\write\lfile{\string\def\string##1\rightbracket}}}}%
\def\writestop{\def\writestoppt{\immediate\write\lfile{\string\pageno
 \the\pageno\string\startrefs\leftbracket\the\refno\rightbracket
 \string\def\string\secsym\leftbracket\secsym\rightbracket
 \string\secno\the\secno\string\meqno\the\meqno}\immediate\closeout\lfile}}
\def\writestoppt{}\def\writedef#1{}

\def\seclab#1{\DefWarn#1%
\xdef #1{\noexpand\hyperref{}{section}{\the\secno}{\the\secno}}%
\writedef{#1\leftbracket#1}\wrlabeL{#1=#1}\par%
\nobreak\medskip\nobreak\noindent\ignorespaces}
\def\subseclab#1\par{\DefWarn#1%
\xdef #1{\noexpand\hyperref{}{subsection}{\the\secno.\the\subsecno}%
{\the\secno.\the\subsecno}}\writedef{#1\leftbracket#1}\wrlabeL{#1=#1}\par%
\nobreak\medskip\nobreak\noindent\ignorespaces}
\def\applab#1{\DefWarn#1%
\xdef #1{\noexpand\hyperref{}{appendix}{\secn@m}{\secn@m}}%
\writedef{#1\leftbracket#1}\wrlabeL{#1=#1}}
\newwrite\tfile \def\writetoca#1{}
\def\leaderfill{\leaders\hbox to 1em{\hss.\hss}\hfill}
\def\writetoc{\immediate\openout\tfile=\jobname.toc
   \def\writetoca##1{{\edef\next{\write\tfile{\noindent ##1
   \string\leaderfill{
   \string\hyperref{}{page}{\noexpand\number\pageno}%
   {\noexpand\number\pageno}} \par}}\next}}
}
\newread\ch@ckfile
\def\listtoc{\immediate\closeout\tfile\immediate\openin\ch@ckfile=\jobname.toc
\ifeof\ch@ckfile\message{no file \jobname.toc, no table of contents this pass}%
\else\closein\ch@ckfile\centerline{\bf Contents}\nobreak\medskip%
{\baselineskip=16pt\footnotefont\parskip=0pt\catcode`\@=11\input\jobname.toc
\catcode`\@=12\bigbreak\bigskip}\fi}
\catcode`\@=12 
\def\tenpoint{\def\rm{\fam0\tenrm}
\textfont0=\tenrm \scriptfont0=\sevenrm \scriptscriptfont0=\fiverm
\textfont1=\teni  \scriptfont1=\seveni  \scriptscriptfont1=\fivei
\textfont2=\tensy \scriptfont2=\sevensy \scriptscriptfont2=\fivesy
\textfont\itfam=\tenit \def\it{\fam\itfam\tenit}\def\footnotefont{\ninepoint}%
\textfont\bffam=\tenbf \def\bf{\fam\bffam\tenbf}\def\sl{\fam\slfam\tensl}\rm}
\font\ninerm=cmr9 \font\sixrm=cmr6 \font\ninei=cmmi9 \font\sixi=cmmi6
\font\ninesy=cmsy9 \font\sixsy=cmsy6 \font\ninebf=cmbx9
\font\nineit=cmti9 \font\ninesl=cmsl9 \skewchar\ninei='177
\skewchar\sixi='177 \skewchar\ninesy='60 \skewchar\sixsy='60
\def\ninepoint{\def\rm{\fam0\ninerm}
\textfont0=\ninerm \scriptfont0=\sixrm \scriptscriptfont0=\fiverm
\textfont1=\ninei \scriptfont1=\sixi \scriptscriptfont1=\fivei
\textfont2=\ninesy \scriptfont2=\sixsy \scriptscriptfont2=\fivesy
\textfont\itfam=\ninei \def\it{\fam\itfam\nineit}\def\sl{\fam\slfam\ninesl}%
\textfont\bffam=\ninebf \def\bf{\fam\bffam\ninebf}\rm}
%
\hyphenation{anom-aly anom-alies coun-ter-term coun-ter-terms}

\global\newcount\subsubsecno \global\subsubsecno=0
\def\subsubsec#1\par{\global\advance\subsubsecno by1%
{\toks0{#1}\message{(\the\secno\the\subsecno\the\subsubsecno. \the\toks0)}}%
\ifnum\lastpenalty>9000\else\bigbreak\fi
\noindent{\it\hyperdef\hypernoname{subsubsection}{\the\secno.\the\subsecno\the\subsubsecno}%
{\the\secno.\the\subsecno.\the\subsubsecno.} #1}
\par\nobreak\medskip\nobreak\noindent\ignorespaces}

\def\DefWarn#1{}
\def\tikzcaption#1#2{\DefWarn#1\xdef#1{Fig.~\the\figno}
\writedef{#1\leftbracket Fig.\noexpand~\the\figno}%
{
\smallskip
\leftskip=20pt \rightskip=20pt \baselineskip12pt\noindent
{{\bf Fig.~\the\figno}\ \ninepoint #2}
\bigskip
\global\advance\figno by1 \par}}

\def\ntoalpha#1{%
\ifcase#1%
@%
\or A\or B\or C\or D\or E\or F\or G\or H\or I
\fi
}

\global\newcount\appno \global\appno=1
\def\applab#1{\xdef #1{\ntoalpha\appno}\writedef{#1\leftbracket#1}\wrlabeL{#1=#1}
\global\advance\appno by1}

\def\preprint#1 #2\par{\rightline{\vbox{\baselineskip12pt\hbox{#1}\hbox{#2}}}\vskip2cm}
%
\def\title#1\par{\centerline{\bf #1}\nopagenumbers\pageno=0}
\def\author#1\par{\bigskip\bigskip\centerline{#1}}

\newcount\addressno

\def\email#1#2{\unskip$^#1$\footnote{\null}{\kern-\parindent \llap{$^#1$\hskip1pt}email: #2}}

\def\startcenter{%
  \par
  \begingroup
  \leftskip=0pt plus 1fil
  \rightskip=\leftskip
  \parindent=0pt
  \parfillskip=0pt
}
\def\stopcenter{\endgroup}

\def\address{\bigskip%
  \ifnum\the\addressno=0\else\stopcenter\endgroup\fi
  \advance\addressno by 1%
  \begingroup
  \startcenter
  \it
  \obeylines
  \addressAux
}
\def\addressAux#1{#1}

\def\abstract{\stopcenter\endgroup\bigskip\bigskip\noindent}

\def\Dsl{\,\raise.15ex\hbox{/}\mkern-13.5mu D} 
\def\dsl{\raise.15ex\hbox{/}\kern-.57em\partial}
\def\tr{{\rm tr}} 
\def\boxeqn#1{\vcenter{\vbox{\hrule\hbox{\vrule\kern3pt\vbox{\kern3pt
	\hbox{${\displaystyle #1}$}\kern3pt}\kern3pt\vrule}\hrule}}}


\def\ap{{\alpha^{\prime}}}

\def\a{\alpha}
\def\b{{\beta}}
\def\g{{\gamma}}
\def\d{{\delta}}
\def\e{{\epsilon}}

\def\s{{\sigma}}

\def\half{{1\over 2}}
\def\p{{\partial}}

\def\({\left(}
\def\){\right)}
\def\dz{{\rm d}z}

\def\cA{{\cal A}}
\def\cF{{\cal F}}


\def\bA{{\Bbb A}}

\def\bF{{\Bbb F}}

\def\Box{\square}
\def\AYM{A^{\rm SYM}}

\def\psum{\mathop{\sum\nolimits'}}

\def\len#1{{%
\def\Dlen{\left|\mkern-1mu #1\mkern -0.5mu\right|}%
\def\Sslen{\left|\mkern-1.3mu #1\mkern -1.3mu\right|}%
\def\SSlen{\left|\mkern-2.8mu #1\mkern-1.3mu\right|}%
\mathchoice{\Dlen}{\Dlen}{\Sslen}{\SSlen}}}

\def\sfrac#1/#2{\kern.1em\raise.5ex\hbox{\the\scriptfont0 #1}%
\kern-.1em/\kern-.15em\lower.25ex\hbox{\the\scriptfont0 #2}}

\font\tenshuffle=shuffle10 \font\sevenshuffle=shuffle7 \font\fiveshuffle=shuffle7 at 5pt
\def\shuffle{{%
\def\Dshuffle{\mathbin{\hbox{\tenshuffle\char'001}}}%
\def\Sshuffle{\mathbin{\hbox{\sevenshuffle\char'001}}}%
\def\SSshuffle{\mathbin{\hbox{\fiveshuffle\char'001}}}%
\mathchoice{\Dshuffle}{\Dshuffle}{\Sshuffle}{\SSshuffle}}}


\def\qed{\hbox{\hskip 3pt
\vbox{\hrule\hbox to 7pt{\vrule height 7pt\hfill\vrule}
\hrule}}\hskip3pt}

\overfullrule=0pt\relax

\frenchspacing

\newread\instream \openin\instream= label.defs
\ifeof\instream \message{No labels in advance yet. Wait till next pass.}
\else \closein\instream \input label.defs
\fi
\writedefs

\def\arXiv:#1].{\hepthStrip#1 \nil}
\def\hepthStrip#1 #2\nil{\href{http://arxiv.org/abs/#1}{arXiv:#1 #2\unskip}].}

\preprint \phantom{M}

\title Berends--Giele recursion for double-color-ordered amplitudes

\author Carlos R. Mafra\email{\star}{mafra@ias.edu}

\address
$^\star$School of Natural Sciences, Institute for Advanced Study,
Einstein Drive, Princeton, NJ 08540, USA

\abstract
Tree-level double-color-ordered amplitudes are computed using Berends--Giele
recursion relations applied to the bi-adjoint cubic scalar theory. The standard
notion of Berends--Giele currents is generalized to double-currents and their
recursions are derived from a perturbiner expansion of linearized fields that solve
the non-linear field equations. Two applications are given. Firstly, we prove that
the entries of the inverse KLT matrix are equal to Berends--Giele double-currents
(and are therefore easy to compute). And secondly, a simple formula to generate
tree-level BCJ-satisfying numerators for arbitrary multiplicity is proposed by
evaluating the field-theory limit of tree-level string amplitudes for various color
orderings using double-color-ordered amplitudes.

\Date {March 2016}


\lref\BerendsME{
	F.A.~Berends and W.T.~Giele,
  	``Recursive Calculations for Processes with n Gluons,''
	Nucl.\ Phys.\ B {\bf 306}, 759 (1988).
}
\lref\PSBCJ{
	C.R.~Mafra, O.~Schlotterer and S.~Stieberger,
	``Explicit BCJ Numerators from Pure Spinors,''
	JHEP {\bf 1107}, 092 (2011).
	[arXiv:1104.5224 [hep-th]].
}
\lref\BGPS{
	C.R.~Mafra and O.~Schlotterer,
  	``Berends-Giele recursions and the BCJ duality in superspace and components,''
	JHEP {\bf 1603}, 097 (2016).
	[arXiv:1510.08846 [hep-th]].
}
\lref\BGSym{
	F.A.~Berends and W.T.~Giele,
	``Multiple Soft Gluon Radiation in Parton Processes,''
	Nucl.\ Phys.\ B {\bf 313}, 595 (1989).
}
\lref\Gauge{
	S.~Lee, C.R.~Mafra and O.~Schlotterer,
  	``Non-linear gauge transformations in $D=10$ SYM theory and the BCJ duality,''
	JHEP {\bf 1603}, 090 (2016).
	[arXiv:1510.08843 [hep-th]].
}
\lref\DPellis{
	F.~Cachazo, S.~He and E.Y.~Yuan,
	``Scattering of Massless Particles: Scalars, Gluons and Gravitons,''
	JHEP {\bf 1407}, 033 (2014).
	[arXiv:1309.0885 [hep-th]].
}
\lref\CHY{
	F.~Cachazo, S.~He and E.Y.~Yuan,
 	``Scattering equations and Kawai-Lewellen-Tye orthogonality,''
	Phys.\ Rev.\ D {\bf 90}, no. 6, 065001 (2014).
	[arXiv:1306.6575 [hep-th]].
}
\lref\dolan{
	L.~Dolan and P.~Goddard,
	``Proof of the Formula of Cachazo, He and Yuan for Yang-Mills Tree Amplitudes in Arbitrary Dimension,''
	JHEP {\bf 1405}, 010 (2014).
	[arXiv:1311.5200 [hep-th]].
}
\lref\KKsym{
	R.~Kleiss and H.~Kuijf,
	``Multi - Gluon Cross-sections and Five Jet Production at Hadron Colliders,''
	Nucl.\ Phys.\ B {\bf 312}, 616 (1989).
}
\lref\KKLance{
	V.~Del Duca, L.J.~Dixon and F.~Maltoni,
	``New color decompositions for gauge amplitudes at tree and loop level,''
	Nucl.\ Phys.\ B {\bf 571}, 51 (2000).
	[hep-ph/9910563].
}
\lref\kkbg{
	C.H.~Fu, Y.J.~Du and B.~Feng,
  	``An algebraic approach to BCJ numerators,''
	JHEP {\bf 1303}, 050 (2013).
	[arXiv:1212.6168 [hep-th]].
}
\lref\oldMomKer{
	Z.~Bern, L.~J.~Dixon, M.~Perelstein and J.~S.~Rozowsky,
	``Multileg one loop gravity amplitudes from gauge theory,''
	Nucl.\ Phys.\ B {\bf 546}, 423 (1999).
	[hep-th/9811140].
}
\lref\MomKer{
	N.~E.~J.~Bjerrum-Bohr, P.~H.~Damgaard, T.~Sondergaard and P.~Vanhove,
	``The Momentum Kernel of Gauge and Gravity Theories,''
	JHEP {\bf 1101}, 001 (2011).
	[arXiv:1010.3933 [hep-th]].
}
\lref\Polylogs{
	J.~Broedel, O.~Schlotterer and S.~Stieberger,
	``Polylogarithms, Multiple Zeta Values and Superstring Amplitudes,''
	Fortsch.\ Phys.\  {\bf 61}, 812 (2013).
	[arXiv:1304.7267 [hep-th]].
}

\lref\nptTreeI{
	C.R.~Mafra, O.~Schlotterer and S.~Stieberger,
	``Complete N-Point Superstring Disk Amplitude I. Pure Spinor Computation,''
	Nucl.\ Phys.\ B {\bf 873}, 419 (2013).
	[arXiv:1106.2645 [hep-th]].
}
\lref\nptTreeII{
	C.R.~Mafra, O.~Schlotterer and S.~Stieberger,
	``Complete N-Point Superstring Disk Amplitude II. Amplitude
	and Hypergeometric Function Structure,''
	Nucl.\ Phys.\ B {\bf 873}, 461 (2013).
	[arXiv:1106.2646 [hep-th]].
}

\lref\nptMethod{
	C.R.~Mafra, O.~Schlotterer, S.~Stieberger and D.~Tsimpis,
	``A recursive method for SYM n-point tree amplitudes,''
	Phys.\ Rev.\ D {\bf 83}, 126012 (2011).
	[arXiv:1012.3981 [hep-th]].
}
\lref\towardsI{
	C.R.~Mafra and O.~Schlotterer,
  	``Towards one-loop SYM amplitudes from the pure spinor BRST cohomology,''
	Fortsch.\ Phys.\  {\bf 63}, no. 2, 105 (2015).
	[arXiv:1410.0668 [hep-th]].
}
\lref\towardsII{
	C.R.~Mafra and O.~Schlotterer,
  	``Two-loop five-point amplitudes of super Yang-Mills and supergravity in pure spinor superspace,''
  	JHEP {\bf 1510}, 124 (2015).
	[arXiv:1505.02746 [hep-th]].
}
\lref\EOMBBs{
	C.R.~Mafra and O.~Schlotterer,
  	``Multiparticle SYM equations of motion and pure spinor BRST blocks,''
	JHEP {\bf 1407}, 153 (2014).
	[arXiv:1404.4986 [hep-th]].
}
\lref\reutenauer{
	C.~Reutenauer,
	``Free Lie Algebras'', London Mathematical Society Monographs, 1993.
}
\lref\Ree{
	R. Ree, ``Lie elements and an algebra associated with shuffles'',
	Ann. Math. {\bf 62}, No. 2 (1958), 210--220.
}
\lref\BGschocker{
	M. Schocker,
	``Lie elements and Knuth relations,'' Canad. J. Math. {\bf 56} (2004), 871-882.
	[math/0209327].
}
\lref\Selivanov{
	K.G.~Selivanov,
	``Postclassicism in tree amplitudes,''
	[hep-th/9905128].
}
\lref\SYM{
	C.R.~Mafra and O.~Schlotterer,
  	``Solution to the nonlinear field equations of ten dimensional
	supersymmetric Yang-Mills theory,''
	Phys.\ Rev.\ D {\bf 92}, no. 6, 066001 (2015).
	[arXiv:1501.05562 [hep-th]].
}
\lref\patras{
	F.~Patras, C.~Reutenauer, M.~Schocker, ``On the Garsia Lie Idempotent'',
   	Canad. Math. Bull. {\bf 48} (2005), 445-454
}
\lref\BaadsgaardVOA{
	C.~Baadsgaard, N.E.J.~Bjerrum-Bohr, J.L.~Bourjaily and P.H.~Damgaard,
  	``Integration Rules for Scattering Equations,''
	JHEP {\bf 1509}, 129 (2015).
	[arXiv:1506.06137 [hep-th]].
}
\lref\BaadsgaardIFA{
	C.~Baadsgaard, N.E.J.~Bjerrum-Bohr, J.L.~Bourjaily and P.H.~Damgaard,
	``Scattering Equations and Feynman Diagrams,''
	JHEP {\bf 1509}, 136 (2015).
	[arXiv:1507.00997 [hep-th]].
}
\lref\LamSQB{
	C.S.~Lam and Y.P.~Yao,
  	``The Role of M\"obius Constants and Scattering Functions in CHY Scalar Amplitudes,''
	[arXiv:1512.05387 [hep-th]].
}
\lref\KLT{
	H.~Kawai, D.C.~Lewellen and S.H.H.~Tye,
	``A Relation Between Tree Amplitudes of Closed and Open Strings,''
	Nucl.\ Phys.\ B {\bf 269}, 1 (1986).
}
\lref\psf{
 	N.~Berkovits,
	``Super-Poincare covariant quantization of the superstring,''
	JHEP {\bf 0004}, 018 (2000)
	[arXiv:hep-th/0001035].
}
\lref\Duhr{
	C.~Duhr, S.~Hoeche and F.~Maltoni,
	``Color-dressed recursive relations for multi-parton amplitudes,''
	JHEP {\bf 0608}, 062 (2006).
	[hep-ph/0607057].
}
\lref\BCJ{
	Z.~Bern, J.J.M.~Carrasco and H.~Johansson,
  	``New Relations for Gauge-Theory Amplitudes,''
	Phys.\ Rev.\ D {\bf 78}, 085011 (2008).
	[arXiv:0805.3993 [hep-ph]].
}
\lref\FORM{
	J.A.M.~Vermaseren,
  	``New features of FORM,''
	[math-ph/0010025].
\semi
	J.~Kuipers, T.~Ueda, J.A.M.~Vermaseren and J.~Vollinga,
	``FORM version 4.0,''
	Comput.\ Phys.\ Commun.\  {\bf 184}, 1453 (2013).
	[arXiv:1203.6543 [cs.SC]].
}


\newsec Introduction

As discussed in \DPellis, the bi-adjoint cubic scalar theory with
the Lagrangian\foot{In \Lagphi\ and
\DPdefCP, $f_{ijk}$ and $\tilde f_{abc}$ are the structure constants of the color
groups $U(N)$ and $U(\tilde N)$ and $t^i$, $\tilde t^a$ are their generators
satisfying $[t^i,t^j] = f^{ijk}t^k$ and $[\tilde t^a,\tilde t^b] = \tilde
f^{abc}\tilde t^c$.}
\eqn\Lagphi{
{\cal L} = \half\p_m \phi_{i|a}\p^m\phi_{i|a}
+ {1\over 3!}f_{ijk}\tilde f_{abc}\phi_{i|a}\phi_{j|b}\phi_{k|c}
}
gives rise to
{\it double-color-ordered} tree amplitudes $m(A|B)$,
\eqn\DPdefCP{
{\cal M}_n = \sum_{a_i,b_i\in S_n/Z_n}\tr(t^{a_1} t^{a_2} \ldots t^{a_n})
\tr(\tilde t^{b_1} \tilde t^{b_2} \ldots \tilde t^{b_n})m(a_1, \ldots,a_n|b_1,
\ldots,b_n),
}
and a diagrammatic algorithm to compute them was described. It was also
demonstrated that these double-color-ordered amplitudes are related to the entries
of the field-theory inverse KLT matrix \refs{\KLT,\oldMomKer,\MomKer} as well as
the field-theory limit of string tree-level integrals
\refs{\nptTreeI,\nptTreeII,\Polylogs}; thus providing an alternative method for
their calculation which does not involve inverting a matrix nor evaluating any
integrals \nptTreeII.

The algorithm to compute $m(A|B)$ described in \DPellis\ involves drawing polygons
and collecting the products of propagators associated to cubic graphs which are
compatible with both color orderings. Their overall sign, however,
requires keeping track of the polygons orientation
in a process that can be challenging to automate. The connection of these
double-color-ordered amplitudes with the Cachazo--He--Yuan approach \CHY\ led to
other recent proposals for their evaluation
\refs{\BaadsgaardVOA,\BaadsgaardIFA,\LamSQB} (see also \dolan).

Given the importance of the double-color-ordered tree amplitudes for the evaluation
of the field-theory limit of string disk integrals, a fully recursive and algebraic
algorithm to compute them will be given in this paper. This will be done using the
{\it perturbiner} approach of \Selivanov\ (recently emphasized in \Gauge) to derive
recursion relations for {\it Berends--Giele double-currents} from a solution to the
non-linear field equation of the action \Lagphi. The double-color-ordered tree
amplitudes are then computed in the same manner as in the Berends--Giele recursive
method \BerendsME.

Two immediate applications of this new method are given. In section~\secInvKLT, the
relation between the inverse field-theory KLT matrix and double-color-ordered
amplitudes observed in \DPellis\ is shown to greatly simplify when the amplitudes
are written in terms of Berends--Giele double-currents. And in section~\secBCJ, the
efficient evaluation of the field-theory limit of string tree-level integrals for
various color orderings will lead to a closed formula for BCJ-satisfying tree-level
numerators \BCJ\ at arbitrary multiplicity, tremendously simplifying the case-by-case
analysis of \PSBCJ.

\subsec On notation

Multiparticle labels correspond to {\it words} in the alphabet
$\{1,2,3,4, \ldots \}$ and are denoted by capital letters (e.g., $A=1243$) while
single-particle labels are represented by lower
case letters (e.g., $i=4$). A word of length $\len{P}$ is
given by $P\equiv p_1p_2 \ldots p_{\len{P}}$ while its transpose is
$\tilde P=p_{\len{P}}p_{\len{P}-1} \ldots p_2p_1$.
The notation $\sum_{XY=P}$ means a sum over all possible ways to
deconcatenate the word $P$ in two non-empty words $X$ and $Y$. For example,
$\sum_{XY=1234} T_X T_Y = T_1 T_{234} + T_{12}T_{34} + T_{123}T_4.$
The shuffle product $\shuffle$ between two words $A$ and $B$
is defined recursively by \reutenauer
\eqn\Shrecurs{
\emptyset\shuffle A = A\shuffle\emptyset = A,\qquad
A\shuffle B \equiv a_1(a_2 \ldots a_{|A|} \shuffle B) + b_1(b_2 \ldots b_{|B|}
\shuffle A)\,,
}
and $\emptyset$ denotes the empty word.
To lighten the notation and avoid summation symbols,
labeled objects are considered to be linear in words; e.g.,
$T_{1\shuffle 23} = T_{123} + T_{213} + T_{231}$.
Finally, the Mandelstam invariants are defined by
\eqn\sdef{
s_P \equiv k_P^2 = (k_{p_1}+ k_{p_2} + \cdots + k_{p_{\len{P}}})^2.
}

\newsec Review of Berends--Giele recursions for Yang--Mills theory

In this section we derive
the Berends--Giele currents for Yang--Mills theory \BerendsME\ from a
solution to the non-linear field equations. This approach has been recently
emphasized in \Gauge\ and resembles the
perturbiner formalism of \Selivanov. The same procedure will be applied in the next
section to the bi-adjoint cubic scalar theory \Lagphi.

The Lagrangian of Yang--Mills theory is given by
\eqn\YMlag{
{\cal L} = -{1\over 4}\tr(\bF_{mn}\bF^{mn}),\qquad \bF_{mn} \equiv -[\nabla_m, \nabla_n]
}
where $\nabla_m = \p_m - \bA_m(x)$ and $\bA_m(x) = \bA_m^a(x) t^a$ is a Lie algebra-valued field
with $t^a$ the generators of a Lie group satisfying $[t^a, t^b]= f^{abc}t^c$. The non-linear
field equation $[\nabla_m, \bF^{mn}] = 0$ following from \YMlag\ can be rewritten in
the Lorenz gauge $\p_m \bA^m = 0$ as
\eqn\EOMYM{
\Box \bA^n(x) = [\bA_m(x),\p^m \bA^n(x)] + [\bA_m(x),\bF^{mn}(x)].
}
To find a solution to the equation \EOMYM\ one writes an ansatz of the
form \refs{\SYM,\Gauge}
\eqn\LieA{
\bA^m(x) \equiv \sum_{P}  \cA^m_{P}(x) t^P,\qquad t^P\equiv t^{p_1}t^{p_2} \ldots
t^{p_{\len{P}}}
}
where the sum is over all words $P$ restricted to permutations.
One can
check using a plane-wave expansion $\cA^m_P(x) = \cA^m_P e^{k_P\cdot x}$
that the ansatz \LieA\ yields the following recursion,
\eqn\cAmrec{
\cA^m_P = -{1\over s_{P}}\sum_{XY=P}\bigl[ \cA^{X}_m (k^X\cdot  \cA^{Y})
+ \cA^{X}_n\cF^Y_{mn} - (X \leftrightarrow Y)\bigr],
}
where $s_P$ is the Mandelstam invariant \sdef,
the field-strength Berends--Giele current is
$\cF^{mn}_Y \equiv k_Y^m \cA_Y^n - k_Y^n \cA_Y^m
- \sum_{RS=Y}\big( \cA_R^m \cA_S^n - \cA_R^n \cA_S^m \big)$
and $\cA^m_i$ with a single-particle label
satisfies the linearized field equation $\Box \cA^m_i = 0$.

It can be shown \BGPS\ that the recursion \cAmrec\ is equivalent to the
recursive definition for the Berends--Giele current $J^m_P$ derived
in \BerendsME\ using Feynman rules for the cubic and quartic vertices of the
Lagrangian \YMlag. Note however that \cAmrec\ contains only ``cubic'' vertices;
the quartic interactions are naturally absorbed by the non-linear terms
of the field-strength. This is conceptually simpler than previous attempts
for absorbing those quartic terms \Duhr.

One can also show using either group-theory methods \BGSym\ or combinatorics of
words \Gauge\ that the currents $\cA^m_P$ satisfy
\eqn\BGsymA{
\cA^m_{A\shuffle B} = 0,\;\; \forall A,B\neq\emptyset\quad \Longleftrightarrow\quad
\cA^m_{PiQ} - (-1)^{\len{P}}\cA^m_{i(\tilde P\shuffle Q)} = 0,
}
which guarantees \Ree\ that the ansatz \LieA\ is a Lie algebra-valued field (the
equivalence between the two statements in \BGsymA\
follows from the theorems proved in \Ree\ and \BGschocker).
Finally, the color-ordered tree-level $n$-point amplitude is given by \BerendsME
\eqn\npttree{
A^{\rm YM}(1,2, \ldots,n) = s_{12 \ldots (n-1)}\cA^m_{12 \ldots(n-1)}\cA^m_n.
}
As a consequence of the Berends--Giele symmetry \BGsymA,
the amplitude \npttree\ {\it manifestly} satisfies the Kleiss--Kuijf symmetry \KKsym;$A^{\rm YM}(P,1,Q,n) = (-1)^{\len{P}}A^{\rm YM}(1,\tilde P\shuffle Q,n)$.
Alternative proofs of this statement are given in \refs{\KKLance,\kkbg}.

\newsec Berends--Giele recursions for the bi-adjoint cubic scalar theory

In this section we derive recursion relations for Berends--Giele double-currents
using a perturbiner expansion for the solution of the non-linear
field equations obtained from the bi-adjoint cubic scalar Lagrangian. These
double-currents will then be used to compute
the tree-level double-color-ordered amplitudes.

\subsec Berends--Giele double-currents

The field equation following from the Lagrangian \Lagphi\ can be written as
\eqn\EOMPhi{
\Box \Phi = [\![\Phi,\Phi]\!]\,,
}
where we defined $\Phi\equiv \phi_{i|a}t^i\tilde t^a$ and
$[\![\Phi,\Phi]\!]\equiv (\phi_{i|a}\phi_{j|b}-\phi_{j|a}\phi_{i|b})t^it^j\tilde
t^a\tilde t^b$.
Following \refs{\SYM,\Gauge},
a solution to the field equation \EOMPhi\ can be constructed
perturbatively in terms of {\it Berends--Giele double-currents} $\phi_{P|Q}$ with
the ansatz,
\eqn\doubleLie{
\Phi(x) \equiv \sum_{P,Q} \phi_{P|Q}\,t^P\tilde t^Q\, e^{k_P\cdot x},
\qquad t^P\equiv t^{p_1}t^{p_2} \ldots t^{p_{|P|}}
}
Since the ansatz \doubleLie\
contains the plane-wave factor $e^{k_P\cdot x}$ (as opposed to $e^{k_Q\cdot x}$),
in order to have a well-defined multiparticle interpretation
$\phi_{P|Q}$ must vanish unless $P$ is a permutation of $Q$, i.e.
$\phi_{P|Q}\equiv 0$ if $P\setminus Q\neq\emptyset$.
Plugging the ansatz \doubleLie\
into the field equation \EOMPhi\
leads to the following recursion
\eqn\BGphi{
\phi_{P|Q} = {1\over s_P}\sum_{XY=P}\sum_{AB=Q}\big(\phi_{X|A}\phi_{Y|B} -
(X\leftrightarrow Y)\big)\,,\quad\hbox{$\phi_{P|Q}\equiv0$, if $P\setminus
Q\neq\emptyset$},
}
where $s_P$ is the multiparticle Mandelstam invariant \sdef\ and the
single-particle double-current\foot{In a slight abuse of notation, the single-particle
double-current $\phi_{i|i}(x)$ in the ansatz \doubleLie\
is not the same field appearing in the Lagrangian \Lagphi; it corresponds
to its linearized truncation.}
satisfies the linearized equation $\Box\phi_{i|i}(x)=0$; therefore $\phi_{i|i}(x)=
\phi_{i|i}e^{k_i\cdot x}$ with $k_i^2=0$ can be normalized such that $\phi_{i|i}=1$.
Since the right-hand side of \BGphi\ is antisymmetric in both
$[XY]$ and $[AB]$,
the combinatorial proof of the Berends--Giele symmetry \BGsymA\
given in the appendix of \Gauge\ also applies to both words in
the double-currents $\phi_{P|Q}$,
\eqn\BGsym{
\phi_{A\shuffle B|Q} = 0  
\quad\Longleftrightarrow\quad
\phi_{AiB|Q} = (-1)^{|A|}\phi_{i(\tilde A\shuffle B)|Q},
}
and, in particular, $\phi_{Ai|Q} = (-1)^{|A|}\phi_{i\tilde A |Q}$ (with
similar expressions for the symmetries w.r.t the word $Q$ in $\phi_{P|Q}$).
The symmetries \BGsym\ generalize the standard Berends--Giele
symmetry \BGsymA\ to both sets of color generators and guarantee that
the ansatz \doubleLie\ is a (double) Lie series \Ree, thereby
preserving the Lie algebra-valued nature of $\Phi(x)$
in \EOMPhi.

Using $\phi_{i|j} = \d_{ij}$
a few example applications of the recursion \BGphi\ are given by
\eqn\simpleEx{
\phi_{12|12} ={1\over s_{12}}\bigl(\phi_{1|1}\phi_{2|2}
-\phi_{2|1}\phi_{1|2})={1\over s_{12}},\quad
\phi_{12|21} ={1\over s_{12}}\bigl(\phi_{1|2}\phi_{2|1} -\phi_{2|2}\phi_{1|1})=-{1\over s_{12}}
}
as well as
\eqnn\moreEx
$$\eqalignno{
\phi_{123|123} &=
 {1\over s_{123}}\bigl(\phi_{12|12} + \phi_{23|23}\bigr) =
{1\over s_{123}}\Bigl({1 \over s_{12}} + { 1 \over s_{23}}\Bigr), &\moreEx\cr
\phi_{123|132} &=
{1\over s_{123}}\phi_{23|32} =- { 1 \over s_{23} s_{123}}.
}$$
In the appendix B, the Berends--Giele double-current $\phi_{P|Q}$
is given an alternative representation in terms of planar binary
trees and products of epsilon tensors.

\subsec Double-color-ordered amplitudes from Berends--Giele double-currents

Without loss of generality, one can use that
$m(R|S)$ is cyclically symmetric
in both words $R$ and $S$ to rewrite an arbitrary $n$-point
amplitude as $m(P,n|Q,n)$, where $|P|=|Q|=n-1$. Therefore,
a straightforward generalization of the gluonic amplitude \npttree\ 
using the Berends--Giele double-currents yields a
formula for the double-color-ordered amplitudes\foot{The
convention for the sign of the Mandelstam invariants here is such that
$m^{\rm here}(P,n|Q,n)= (-1)^{|P|}m^{\rm there}(P,n|Q,n)$ in
comparison with the normalization of \DPellis.} (recall that
$\phi_{n|n}=1$),
\eqn\BGamplitude{
m(P,n|Q,n) = s_P \phi_{P|Q}\,.
}
It is easy to see using
the symmetries \BGsym\ obeyed by the double-currents that
the Kleiss--Kuijf relations are satisfied independently by both sets of color
orderings.
Since the double-currents $\phi_{P|Q}$ obey
the recursion relation \BGphi, the computation of double-color-ordered
amplitudes is easy to automate and their overall
sign requires no additional bookkeeping\foot{An implementation using FORM \FORM\ is
attached to the arXiv submission.}.

\newsec The field-theory KLT matrix and its inverse
\par\seclab\secInvKLT

\noindent In this section we demonstrate that the entries of the
inverse field-theory KLT matrix \refs{\KLT,\oldMomKer} (also
called the {\it momentum kernel matrix} \MomKer)
are equal to the Berends--Giele double currents and therefore are
easy to compute. This computational simplicity is important because,
apart from applications related to gauge and gravity amplitudes, the
field-theory KLT matrix
and its inverse relate \Polylogs\ the local and non-local versions
of multiparticle super Yang--Mills superfields\foot{The relations \VtoM\
apply for all types of SYM superfields ($A_\a, A_m, W^\a, \ldots$) \EOMBBs. The
restriction to $V_P$ in \VtoM\ was chosen for simplicity.}
\eqn\VtoM{
M_{1A} = \sum_B S^{-1}[A|B]_1 V_{1B},\qquad
V_{1A} = \sum_B S[A|B]_1 M_{1B},
}
with manifold applications in recent developments within
the pure spinor formalism applied to the computation of scattering
amplitudes in both field- and string theory \refs{\nptTreeI,\nptMethod,\towardsI,\towardsII}.

\subsec The field-theory KLT matrix

The symmetric matrix $S[P|Q]$ defined by
\eqn\momKdef{
S[P|q_1q_2 \ldots q_\len{Q}] \equiv
\prod_{j=2}^{\len{Q}}\sum_{i=1}^{j-1} s(P|q_i,q_j),\quad 
s(P|q_i,q_j) \equiv \cases{s_{q_iq_j}, & $q_i<q_j$ inside $P$ \cr 0, & otherwise}
}
gives rise to the KLT matrix $S[A|B]_i$
when the first letters on both words coincide
\eqn\Sdef{
S[A|B]_i \equiv S(i,A|i,B)\,.
}
For example, the definition \Sdef\ for $i=1$ leads to
$S[2|2]_1 = s_{12}$ as well as
$$\eqalign{
S[23|23]_1 & = s_{12}(s_{13}+s_{23}),\cr
S[234|234]_1 &= s_{12} (s_{13}+s_{23}) (s_{14}+s_{24}+s_{34}),\cr
S[243|234]_1 &= s_{12} (s_{13}+s_{23}) (s_{14}+s_{24}),\cr
S[324|234]_1 &= s_{12} s_{13} (s_{14}+s_{24}+s_{34}),\cr
}\qquad\eqalign{
S[23|32]_1 & = s_{12}s_{13},\cr
S[423|234]_1 &= s_{12} (s_{13}+s_{23}) s_{14},\cr
S[342|234]_1 &= s_{12} s_{13} (s_{14}+s_{34}),\cr
S[432|234]_1 &= s_{12} s_{13} s_{14}.\cr
}$$

\subsec The inverse KLT matrix

The inverse KLT matrix $S^{-1}[A|B]_i$ can be computed from the entries \Sdef\ 
using standard matrix algebra. However, this task
quickly becomes tedious in practice and the direct outcome of the matrix inversion
usually requires further manipulations to be simplified.
Fortunately it  was proven in \DPellis\ that the entries of $S^{-1}[A|B]_i$
correspond to the double-color-ordered amplitudes\foot{The overall sign in
\InvSDPa\ is different than in \DPellis\ due to differences in conventions.},
\eqn\InvSDPa{
S^{-1}[A|B]_i = - m(i,A,n-1,n|i,B,n,n-1),\quad |A|=|B|=n-3,
}
completely bypassing the tedious matrix algebra necessary to invert the
KLT matrix \momKdef.
With the Berends--Giele representation of double-color-ordered
amplitudes \BGamplitude\
the computation of $S^{-1}[A|B]_i$ does not require the extra labels $n-1,n$
since \InvSDPa\ simplifies to
\eqn\InvBG{
S^{-1}[A|B]_i = \phi_{iA|iB}.
}
To see this one uses the Berends--Giele amplitude formula \BGamplitude\ in \InvSDPa\ to obtain
\eqnn\prooftmpI
$$\eqalignno{
S^{-1}[A|B]_i &= - s_{iA(n-1)}\phi_{iA(n-1)|(n-1)iB} =
(-1)^{\len{A}}s_{iA(n-1)}\phi_{(n-1)\tilde A i|(n-1)iB} &\prooftmpI\cr
&= (-1)^{|A|} \phi_{(n-1)|(n-1)}\phi_{\tilde A i|iB} =\phi_{iA|iB}\,.
}$$
In the first line the label $(n-1)$ has been moved to the
front using \BGsym
\eqn\symmove{
\phi_{iA(n-1)|P} = (-1)^{\len{A}+1}\phi_{(n-1)(\widetilde{iA})|P} = -
(-1)^{\len{A}}\phi_{(n-1)\tilde Ai|P},
}
and in the second line the condition $\phi_{P|Q}=0$ unless $P$ is a permutation of $Q$
implies that
$s_{iA(n-1)} \phi_{(n-1)\tilde A i|(n-1)iB} = \phi_{(n-1)|(n-1)}\phi_{\tilde A
i|iB}$.
For example,
\eqnn\Sexamp
$$\eqalignno{
S^{-1}[23|23]_1 &= \phi_{123|123} = {1 \over s_{12} s_{123}} + { 1 \over s_{23} s_{123}},
\quad
S^{-1}[23|32]_1 = \phi_{123|132} = - { 1 \over s_{23} s_{123}}\,,\cr
S^{-1}[32|32]_1 &= \phi_{132|132} = {1 \over s_{13} s_{123}} + { 1 \over s_{23} s_{123}},
&\Sexamp
}$$
which agrees with the results of \Polylogs. Higher-multiplicity
examples follow similarly.

Using the Berends--Giele representation of the inverse KLT matrix \InvBG,
the first relation in \InvSDPa\ simplifies to
\eqn\simpleVtoM{
M_{1A}=\sum_B \phi_{1A|1B}V_{1B},
}
and therefore provides an efficient algebraic alternative to
the diagrammatic method to compute $M_P$ described in the appendix of \EOMBBs.

\newsec The field-theory limit of tree-level string integrals

The $n$-point open-string amplitude computed using pure spinor methods in
\nptTreeI\ can be written in terms of (local) multiparticle vertex operators $V_P$ \EOMBBs\ as
\eqn\stringtree{
A(\Sigma) = \psum_{XY=2 \ldots n-2}\!\!\!\!\! \langle V_{1X} V_{(n-1)\tilde Y}V_n\rangle
Z_\Sigma(1,X,n,Y,n-1)(-1)^\len{X} + {\cal P}(23 \ldots n-2),
}
where the deconcatenation in $\psum_{XY}$ includes empty words and
$Z_\Sigma(N)$ is given by \Polylogs,
\eqn\Zdef{
Z_\Sigma(1,2,3, \ldots n-1,n) \equiv {1\over {\rm vol}(SL(2,\Bbb R))}
\int_{\Sigma}\! \dz_1\dz_2\cdots \dz_n
{\prod_{i<j}^n |z_{ij}|^{\ap s_{ij}}\over z_{12}z_{23}\cdots z_{n-1,n} z_{n1}}.
}
The factor $1/{\rm vol}(SL(2,\Bbb R))$ compensates
the overcounting due to the conformal Killing group of the disk\foot{It
amounts to fixing three coordinates $z_i,z_j$
and $z_k$ and inserting a Jacobian factor $|z_{ij}z_{jk}z_{ki}|$.} and the region
of integration $\Sigma$ is such that $z_{\s_i}< z_{\s_{i+1}}$ for all $i=1$ to
$i=\len{M}-1$. The pure spinor bracket $\langle \ldots\rangle$ is defined in \psf\
but will play no role in the subsequent discussion.

As pointed out in \DPellis, the field-theory limit of the string disk
integrals \Zdef\ is given by the double-color-ordered amplitudes,
\eqn\apzero{
\lim_{\ap\to0}Z_P(Q) = (-1)^{\len{P}}m(P|Q)\,.
}
For example ($I=123 \ldots n$),
\eqnn\FTex
$$\eqalignno{
\lim_{\ap\to0}Z_I(1243) &= (-1)^4m(1234|1243) = s_{123}\phi_{123|312} = - {1\over s_{12}}
&\FTex\cr
\lim_{\ap\to0}Z_I(12354) &= (-1)^5m(12345|12354) = -s_{1234}\phi_{1234|4123} =
{1\over s_{12}s_{123}}
+ {1\over s_{23}s_{123}},
}$$
which agree with (C.1) and (C.5) of \Polylogs. Higher-multiplicity examples follow
similarly.

So the SYM tree amplitudes with color ordering $\Sigma$
obtained from the field-theory limit of the string amplitude
\stringtree\ are given by
\eqn\SYMstring{
A^{\rm SYM}(\Sigma) =\!\!\!\! \psum_{XY=2 \ldots n-2}\!\!\!\!\! \langle V_{1X}
V_{(n-1)\tilde Y}V_n\rangle\,
m(\Sigma|1,X,n,Y,n-1)(-1)^{\len{Y}+1} + {\cal P}(23 \ldots n-2).
}
It was shown in \PSBCJ\ that a set of BCJ-satisfying numerators
for SYM tree amplitudes can always be obtained from
the field-theory limit of the string tree-level amplitude \stringtree,
and explicit expressions for numerators up to $7$-points were given
in that reference.
Since the Berends--Giele
algorithm to evaluate the double-color-ordered amplitudes is easy to automate,
one can quickly obtain higher-point BCJ numerators this way.
Studying their patterns leads to a proposal
for a general formula giving BCJ-satisfying tree-level
numerators for arbitrary multiplicities. This will be done in the next section.

\newsec Tree-level SYM amplitudes with manifest BCJ numerators
\par\seclab\secBCJ

\noindent For the canonical ordering $\Sigma=123 \ldots n$ it is easy to see that
\SYMstring\ reproduces the pure spinor $n$-point SYM amplitude formula derived in \nptMethod\
\eqn\SYMtree{
A^{\rm SYM}(1,2, \ldots,n) = \langle E_{12 \ldots n-1}V_n\rangle,\qquad
E_P \equiv \sum_{XY=P}M_XM_Y\,,
}
where $M_X$ denotes the Berends--Giele current \simpleVtoM\ associated with the multiparticle
vertex $V_X$ \EOMBBs.
As discussed in \PSBCJ, the amplitude \SYMtree\
is the supersymmetric generalization of
the standard Berends--Giele recursions \BerendsME\ and leads to an
alternative proof of the Kleiss--Kuijf relations \KKsym\
(originally proven in \KKLance).

To prove that \SYMstring\ reduces to \SYMtree\ when $\Sigma=123 \ldots n$,
note that $m(\Sigma|1,X,n,Y,n-1)$ simplifies
when $X$ and $Y$ are also canonically ordered (which is the case for \SYMstring),
\eqn\msimple{
m(12 \ldots n|1,X,n,Y,n-1) = s_{12 \ldots n-1}\phi_{12 \ldots n-1|Y(n-1)1X}
= - \phi_{1X|1X}\phi_{Y(n-1)|Y(n-1)}.
}
Therefore the field-theory limit of the string tree amplitude given in
\SYMstring\ becomes
\eqnn\SYMstringFT
$$\eqalignno{
A^{\rm SYM}(12 \ldots n) &=\!\!\! \psum_{XY=2 \ldots n-2}\!\!\!\!\! \langle V_{1X}
V_{(n-1)\tilde Y}V_n\rangle\, \phi_{1X|1X}\phi_{Y(n-1)|Y(n-1)}
(-1)^{\len{Y}} + {\cal P}(23 \ldots n-2)\cr
&=\!\!\!\!\!\psum_{XY=2 \ldots n-2}\!\!\!\!\! \langle M_{1X}M_{Y(n-1)}V_n\rangle =
\!\!\!\!\!\sum_{XY=1 \ldots n-1}\!\!\!\!\!\langle M_X M_Y V_n\rangle =
\langle E_{12 \ldots n-1}V_n\rangle, &\SYMstringFT
}$$
where $\phi_{Y(n-1)|Y(n-1)} =
\phi_{(n-1)\tilde Y|(n-1)\tilde Y}$ was used before applying
\simpleVtoM\ to identify
$M_{1X} = \sum_{P} \phi_{1X|1P}V_{1P}$ and
$M_{(n-1)\tilde Y} = \sum_{P} \phi_{(n-1)\tilde Y|(n-1)P}V_{(n-1)P} =
(-1)^\len{Y}M_{Y(n-1)}$. Note that the permutations over $23 \ldots n-2$
do not act on the labels corresponding to the canonical ordering in $\phi_{1X|1X}$
such that $\phi_{1X|1X}V_{1X} + {\cal P}(23 \ldots n-2) = \sum_P \phi_{1X|1P}V_{1P}$.

However, for general color orderings \SYMstring\ and \SYMtree\
no longer manifestly coincide. For example, the field-theory limit
of the string amplitude \SYMstring\
with ordering $12435$ is
\eqnn\patta
$$\eqalignno{
\AYM(1,2,4,3,5) &=
{\langle(V_{12}V_{43} + V_{123}V_4)V_5\rangle \over s_{12}s_{124}}
- { \langle(V_1 V_{423} + V_{13}V_{42})V_5\rangle \over s_{24}s_{124}}
+ {\langle V_{12}V_{43}V_5\rangle\over s_{34}s_{12}} \cr
&- { \langle V_1V_{432}V_5\rangle\over s_{34}s_{234}}
- {\langle V_1V_{423}V_5\rangle\over s_{24}s_{234}}\,,&\patta
}$$
while the field-theory formula \SYMtree\ yields
\eqnn\Fivetwo
$$\eqalignno{
\AYM(1,2,4,3,5) &= \langle E_{1243}M_5\rangle
= \langle \big(M_{124}M_3 + M_{12}M_{43} + M_1 M_{243}\big)M_5\rangle &\Fivetwo\cr
& = {\langle V_{124}V_3V_5\rangle\over s_{12}s_{124}}
+ {\langle V_{421}V_3V_5\rangle\over s_{24}s_{124}}
+ {\langle V_{12}V_{43}V_5\rangle\over s_{12}s_{34}}
+ {\langle V_1V_{243}V_5\rangle\over s_{24} s_{34}}
+ {\langle V_1V_{342}V_5\rangle\over s_{34}s_{234}}\,.
}$$
One can see from \Fivetwo\ and \patta\ that the numerators generated by
the SYM amplitude formula \SYMtree\ are mapped to the following
BCJ-satisfying
numerators in the string theory amplitude,
\eqn\strmod{\eqalign{
V_{124}V_3 &\rightarrow V_{12}V_{43} + V_{123}V_4,\cr
V_{421}V_3 &\rightarrow - V_1 V_{423} - V_{13}V_{42},
}\quad\eqalign{
V_1V_{243} &\rightarrow -V_1V_{423},\cr
V_1V_{342} &\rightarrow -V_1V_{432}.
}}
Comparing the field-theory limit of the string
amplitude \SYMstring\ for various orderings with the outcomes of
the SYM amplitude \SYMtree, one can check that the BCJ-satisfying numerators
following from the string tree amplitude
can be obtained by a mapping $\circ_{ij}$ defined by
\eqnn\Prodij
$$\eqalignno{
V_{iAjB}\circ_{ij} V_C &\equiv \sum_{\a\in P(\g)} V_{iA\a}V_{j\b},\quad
\g \equiv\{ B, \ell(C)\},\quad \b \equiv \g\backslash \a, &\Prodij\cr
V_{AiB}\circ_{ij} V_{CjD} &\equiv V_{AiB}V_{CjD}
}$$
acting\foot{It suffices to define $\circ_{ij}$ as in \Prodij\ since the
generalized Jacobi identity $V_{AiB} = - V_{i\ell(A)B}$ \Gauge\ can
always be used to move the label $i$ to the front.}
on the field-theory numerators given by the SYM amplitude \SYMtree. In \Prodij,
$P(\g)$ denotes the powerset of $\g$, $\ell(C)$ is the left-to-right
Dynkin bracket \reutenauer,
\eqn\dynkin{
\ell(c_1c_2c_3 \ldots c_\len{C}) \equiv [[ \ldots[c_1,c_2],c_3], \ldots,],c_\len{C}]
}
and $\ell(C)$ is considered a single letter in the definition
of the powerset of $\g=\{B,\ell(C)\}$; the number of elements in $P(\g)$ is
$2^{\len{B}+1}$.

The mapping \Prodij\ ensures that the labels $i$ and $j$
never belong to the same vertex $V_A$ or $V_B$ in
the product $V_A \circ_{ij}V_B$. This corresponds to the label distribution in the
string theory formula \SYMstringFT\ if $i=1$ and $j=n-1$ and is the result
of fixing the M\"obius symmetry of the disk.
For example, in a five-point amplitude one chooses
$i=1$ and $j=4$ to get,
\eqnn\mobex
$$\eqalignno{
V_{124}\circ_{14}V_3 &= V_{12} V_{43} +  V_{123} V_{4} &\mobex\cr
V_{142}\circ_{14}V_3 &= V_{1} V_{423} +  V_{12} V_{43} + V_{123}V_4 + V_{13}V_{42}\cr
V_{421}\circ_{14}V_3 &= -  V_{1} V_{423} -  V_{13} V_{42}\cr
%
%
}$$
Defining $M_X\circ_{ij} M_Y$ by its action
on the products of $V_A\circ_{ij} V_B$ from the expansion of $M_X$ and
$M_Y$ given by \simpleVtoM\ one can check a few cases explicitly that
the following superfield is BRST closed
($Q$ is the pure spinor BRST charge \psf)
\eqn\Eijdef{
E_P^{(ij)}\equiv \sum_{XY=P} M_X\circ_{ij} M_Y \quad\Longrightarrow\quad
QE^{(ij)}_P = 0,\quad \forall i,j \in P.
}
Assuming that $E^{(ij)}_P$ is BRST invariant to all multiplicities,
one is free to use this ``gauge-fixed'' version of $E_P$ in the
SYM amplitude formula \SYMtree\ to obtain
\eqn\BCJamps{
\AYM(1,2,3, \ldots,n) \equiv \langle E^{(ij)}_{123 \ldots n-1}V_n\rangle,\qquad
i,j\equiv (1,n-1)\,.
}
By construction, the SYM amplitudes generated by the formula \BCJamps\ manifestly coincide
with the field-theory limit of the string tree amplitude and therefore give
rise to BCJ-satisfying numerators for all $n$-point tree amplitudes. Incidentally,
the powerset appearing in the definition \Prodij\ naturally explains why
the number of terms in BCJ-satisfying numerators is always a power of two,
as firstly observed in \PSBCJ.

In the appendix~\appBCJ\
the mapping \Prodij\ is shown to be the kinematic equivalent of the color
Jacobi identity which expresses
any cubic color graph in a basis where labels $i$ and $j$ are
at the opposite ends.

\bigskip
\noindent{\bf Acknowledgements:} I thank Ellis Yuan for discussions about the
algorithm of \DPellis\ and Oliver Schlotterer for
an enlightening discussion about the perturbiner expansion of Berends--Giele
double-currents as well as for collaboration on related topics.
I also acknowledge support
from NSF grant number PHY 1314311 and the Paul Dirac Fund.

\appendix{A}{Proof of manifest BCJ numerators}
\applab\appBCJ

\noindent In this appendix we prove that the rewriting of
field-theory numerators given by \Prodij\ corresponds to the Jacobi identity
obeyed by structure constants.

In a BCJ gauge of super Yang--Mills superfields,
the multiparticle vertex operator $V_P$ satisfies generalized Jacobi identities
(see e.g. \reutenauer) and therefore its symmetries correspond to a string of
structure constants \EOMBBs
\eqn\LieBasis{
V_{AiB} = - V_{i\ell(A)B}\quad \Longleftrightarrow\quad
V_{1234 \ldots p }\leftrightarrow f^{12a_3}f^{a_3 3 a_4}f^{a_4 4 a_5} \cdots f^{a_p
p a_{p+1}},
}
where $\ell(A)$ denotes the Dynkin bracket \dynkin.
Similarly, the symmetries of three vertices are mapped to
\eqn\VVVnum{
V_{iAjB} V_C V_n \Longleftrightarrow (-1)^{|C|}F(i,A,j,B,n,\tilde C)\,,
}
where $F(A)$ is the {\it multi-peripheral} color factor \KKLance\ 
\eqn\Fcolor{
F(1,2,3, \ldots,(n-1),n)\equiv
f^{12 a_3} f^{a_33a_4} f^{a_44a_5}\cdots f^{a_{(n-1)}(n-1)n}.
}
Applying the generalized Jacobi identity \LieBasis\ either once or twice,
any multi-peripheral color factor can
be rewritten in the Del Duca--Dixon--Maltoni (DDM) basis of \KKLance
\eqn\cbasisKK{
F(A,i,B,j,C) = \cases{
F(i,\ell(A),B,\tilde \ell(\tilde C),j), & $A\neq\emptyset$, $C\neq\emptyset$\cr
-F(i,B,\tilde \ell(\tilde C),j), & $A =\emptyset$, $C\neq\emptyset$ \cr
}
}
where $\tilde\ell(P)=\widetilde{\ell(P)}$.
One can also derive a closed formula to arrive at
the DDM basis while keeping track of the relative positions of
three labels (say $i$, $j$ and $n$),
\eqnn\cKKid
$$\eqalignno{
F(i,A,j,B,n,C)&= - F(i,A,\tilde \ell(\tilde C n\tilde B),j),\quad C\neq\emptyset &\cKKid\cr
& =\sum_{\a\in P(\g)} (-1)^{\len{\b}}F(i,A,\tilde\a,n,\b,j),
\quad \g\equiv\{\ell(\tilde C),\tilde B\},\;\; \b\equiv\g\backslash\a
}$$
where $P(\g)$ is the powerset of $\g$ and $\ell(\tilde C)$ is to be 
considered a single letter in $P(\g)$. To arrive at the
second line one uses the identity\foot{When  $P=\emptyset$, the sign
factor is given by $(-1)^{\len{\b}}$.} (see \patras)
\eqn\dynid{
\ell(PnQ) = \sum_{\a\in P(\g)}(-1)^{\len{\b}+1}\tilde\b\, n\,\a,\quad
P\neq\emptyset,\;\; \g\equiv\{\ell(P),Q\},\;\; \b\equiv\g\backslash\a
}
Finally, combining the results above one gets
\eqnn\jacproof
$$\eqalignno{
V_{iAjB}V_C V_n &\rightarrow (-1)^{\len{C}}F(i,A,j,B,n,\tilde C) &\jacproof\cr
&=(-1)^\len{C}\!\!\!\sum_{\a\in P(\d)} (-1)^{\len{\b}}F(i,A,\tilde\a,n,\b,j),
\quad \d\equiv\{\ell(C),\tilde B\},\;\;\b\equiv\d\backslash\a\cr
&\rightarrow (-1)^{\len{C}+1}\!\!\!\sum_{\a\in P(\d)}
V_{iA\tilde\a}V_{j\tilde\b} V_n\cr
&=\sum_{\a\in P(\g)}
V_{iA\a}V_{j\b} V_n,
\quad \g\equiv\{B,\ell(C)\},\;\;\b\equiv\g\backslash\a\cr
&= V_{iAjB}\circ_{ij}V_C V_n
}$$
where in the penultimate line we transposed the set $\d$ (while considering
$\ell(C)$ a single letter) and used
$\tilde \ell(C) = (-1)^{\len{C}+1}\ell(C)$ when $\ell(C)$ is part of a multiparticle label.

Therefore the expression \Prodij\ for the product $V_{iAjB}\circ_{ij}V_C V_n$ is the
kinematic counterpart of the color identity \cKKid.

\appendix{B}{Berends--Giele double-currents from scalar $\phi^3$ theory}

\noindent In this appendix an alternative derivation of the Berends--Giele
double-currents is given which resembles the algorithm of \DPellis.

The field equation $\Box\phi=\phi^2$ of the standard scalar $\phi^3$ theory
can be solved in a perturbiner expansion as $\phi(x)=\sum_P \phi_P e^{k\cdot x}
\xi^P$, where $\xi^P = \xi^{p_1}\xi^{p_2} \ldots \xi^{p_{\len{P}}}$ is an auxiliary
parameter and the coefficients $\phi_P$ obey the recursion relations of planar
binary trees,
\eqn\BGtree{
\phi_i = 1,\qquad \phi_P = {1\over s_P}\sum_{XY=P}\phi_X \phi_Y\,,\quad X,Y\neq\emptyset.
}
It is straightforward to check
that \BGtree\ gives rise to the recurrence relation for the Catalan numbers,
$C_0=1$, $C_{n+1} =\sum_{i=0}^n C_i C_{n-i}$, where $C_n$ refers to the number
of terms in the pole expansion of $\phi_{12 \ldots n+1}$.
Examples of $\phi_{123 \ldots n}$ up to $n=4$ are given by,
\eqnn\bintrees
$$\displaylines{
\phi_1 = 1,\quad\quad \phi_{12} = {1\over s_{12}}, \quad \quad
\phi_{123} = {1 \over s_{12} s_{123}} + { 1 \over s_{23}s_{123}}\,,\hfil\bintrees\hfilneg\cr
\phi_{1234} = {1 \over s_{1234}} \Big( {1\over s_{12}s_{123} }
 + {1\over s_{23}s_{123} } + {1 \over s_{12}s_{34} }
 + {1 \over s_{34}s_{234} } + {1 \over s_{23}s_{234} } \Big)\,.
}$$
Note that the above binary trees naturally capture the kinematic pole expansion of
``compatible channels'' in a color-ordered tree amplitude.

The restriction of $\phi_P$ by an ordering given by a
word $A$ is denoted $\phi_P\big|_A$ and is
defined by suppressing a term from $\phi_P$ if
it contains any factor of $s_{abcd \ldots}$ whose letters
are not adjacent in the word $A$.
For example, if $A=1324$ then
\eqn\exaOrder{
A=1324 \Longrightarrow
\cases{s_{13},s_{23},s_{24},s_{123},s_{234},s_{1234}
& allowed\cr
s_{12},s_{14},s_{34},s_{124},s_{134} & not allowed}
}
and the restriction of $\phi_{1234}$ by $A=1324$ yields
\eqn\phiEx{
\phi_{1234}\big|_{1324} = {1 \over s_{1234}} \Big(
{1\over s_{23}s_{123} } + {1 \over s_{23}s_{234} } \Big)\,.
}
Now define a sign factor as follows
\eqn\eABdef{
\e_{A|B} \equiv \e(A|b_1,b_2)\e(A|b_2,b_3)
\ldots \e(A|b_{p-1},b_p),\quad
\e(A|i,j) \equiv \cases{+ 1, & $i<j$ inside $A$ \cr -1, & $i>j$ inside $A$}
}
where ``$i<j$ inside $A$'' is true if the letter $i$ appears before $j$ in $A$.
For example, $\e(1324|1,4) = +1$ but $\e(1324|4,1)=-1$.
If $P= 123\ldots p$ is the canonical ordering, the sign factor simplifies to
$\e(P|Q) = \e_{q_1q_2}\e_{q_2q_3} \ldots \e_{q_{\len{Q}-1}q_\len{Q}}$ where
$\e_{ij}$ is the standard anti-symmetric tensor;
$\e_{ij} = + 1$ if $i<j$ and $\e_{ij}= -1$ if $i>j$.

One can check that the Berends--Giele double-currents \BGphi\ can be written as
\eqn\dBGPBT{
\phi_{P|Q} \equiv \e(P|Q)\phi_P\big|_Q.
}
Comparing \dBGPBT\ with the algorithm of \DPellis\ one concludes that
the cumbersome factor of $(-1)^{n_{\rm flip}}$ of \DPellis\ admits
a simpler representation in terms of epsilon tensors (this observation
was made {\it en passant} in \LamSQB).

\ninerm
\listrefs
\bye